# Temperature dependence of near-field radiative heat transfer above room temperature

C. Lucchesi[1,2]*, R. Vaillon[2,1], P.-O. Chapuis[1]

[1] Univ Lyon, CNRS, INSA-Lyon, Université Claude Bernard Lyon 1, CETHIL UMR5008, F-69621, Villeurbanne, France
[2] IES, Univ Montpellier, CNRS, Montpellier, France

christophe.lucchesi@insa-lyon.fr

**Abstract**
Stefan-Boltzmann's law indicates that far-field blackbody radiation scales at the fourth power of temperature. The temperature dependence of radiative heat transfer in the near field is expected to be very different due to the contribution of evanescent waves. In this work, we experimentally observe such deviation on the radiative thermal conductance by bringing a hot micrometric sphere in the near-field of a room-temperature planar substrate, down to a separation distance of few tens of nanometers. The influence of materials is assessed by using either $SiO_2$ or graphite spheres, and $SiO_2$, graphite and InSb substrates. Temperature differences as large as 900 K are imposed. A maximum near-field radiative thermal conductance of about 70 nW.K$^{-1}$ is found for a graphite-graphite configuration. The experimental results demonstrate that the temperature exponent weakens in the near field, ranging from 2.2 to 4.1, depending on the gap distance and the materials. These results have broad consequences, in particular on the design of high-temperature nanoscale radiative energy devices.

## 1. Introduction

Physics of radiative heat transfer at the nanoscale is very different from that at the macroscale. The classical theory of thermal radiation fails to describe radiative heat transfer when the distance separating two bodies is smaller than the characteristic wavelength of thermal radiation[1] ($\lambda_{Wien}$~10 µm at room temperature and ~2.3 µm near 1000 °C). The range corresponding to such distances is referred to as the near-field regime, while the far-field regime corresponds to the macroscale theory involving propagative waves. In the near field, a new path for thermal radiation emerges due to the contribution of the evanescent waves, known as photon tunneling. The contribution of these waves was theoretically predicted and experimentally confirmed to enhance the thermal radiative power exchanged between two bodies in the near field by up to several orders of magnitude compared to the far field[1–4]. The first attempt to demonstrate such an enhancement was performed by Hargeaves[5] in 1969. From then on, many experimental proofs of the near-field enhancement of radiative heat transfer were reported with various geometries (plane[6–9], sphere[3,10,11] or tip[4,12,13]) and materials (see Lucchesi *et al.*[14] for an exhaustive review). Many experimental works[2,3,7,10,11,15–31] used $SiO_2$ as the material of the two bodies, leading to a large enhancement of near-field radiative heat transfer as this material supports surface phonon polaritons[32]. In addition, only relatively small temperature differences with $\Delta T < 425$ K were investigated experimentally[7]. It is striking that almost all works investigated only the distance dependence of radiative heat transfer, while the temperature dependence of the exchanged power could also be modified. Indeed, the dimensionless parameter $x = hc/\lambda k_B T$ in Planck's law is approximately replaced by $hc/dk_B T$ in the expression of the near-field radiative flux, which underlines a key coupling between distance and temperature[33]. A deviation from the well-known fourth power of temperature is therefore expected.

An early analysis at low temperature was performed in the micrometric regime[6]. The temperature dependence was also analyzed theoretically between planar surfaces[33]. It was also shown that emitters of sub-wavelength size can radiate with temperature power laws that deviate from the Stefan-Boltzmann's prediction[34–37]. In this work, we study experimentally the temperature dependence of near-field radiative heat transfer between two bodies separated by distances down to few tens of nanometers and for hundreds of kelvins above room temperature, a range which is key to many fields including high-temperature energy-conversion devices (thermophotovoltaics, thermoradiative, etc.). Near-field radiative measurements between a spherical emitter and a planar substrate are therefore implemented for large temperature differences up to 900 K.

## 2. Methods

### 2.1. Experimental setup

The emitter is a sphere made either of $SiO_2$ or graphite with a diameter respectively equal to 44 or 37.5 µm. It is glued on the tip of a doped-silicon scanning thermal microscopy (SThM) probe by using an alumina-based ceramic adhesive supposed to withstand high temperatures up to 1900 K. The sphere is heated by heat conduction occurring at the apex of the cantilever, which is self-heated when an electrical current is applied (Fig. 1a). The temperature dependence of the electrical resistance of the probe (Fig. 1b, see section 2.2 for the calibration) allows inferring the temperature at the apex, considered to be that of the sphere. Note that thermal steady state is reached in the {cantilever+sphere system} after a time of about 30 min. The temperature change of the emitter as the gap distance with the substrate is reduced allows determining the increase of thermal conductance due to near-field effects. Such a calibration curve is key for a proper analysis of the experimental data and is detailed elsewhere[38–40]. Near-field radiative heat transfer experiments are performed by placing the probe-sphere assembly on a 3-axis piezoelectric positioning system located in a vacuum chamber in which the pressure is maintained at ~10⁻⁶ mbar by an ion pump for avoiding any mechanical vibration. The emitter is positioned at a distance of about 5 µm above the substrate. The z-piezoelectric actuator displacement step is set to 1.7 nm. The temperature of the emitter is measured while a voltage is applied to the z-piezoelectric positioner that expands until the sphere is in contact with the substrate. The contact is determined as it induces a sharp temperature drop caused by thermal conduction.

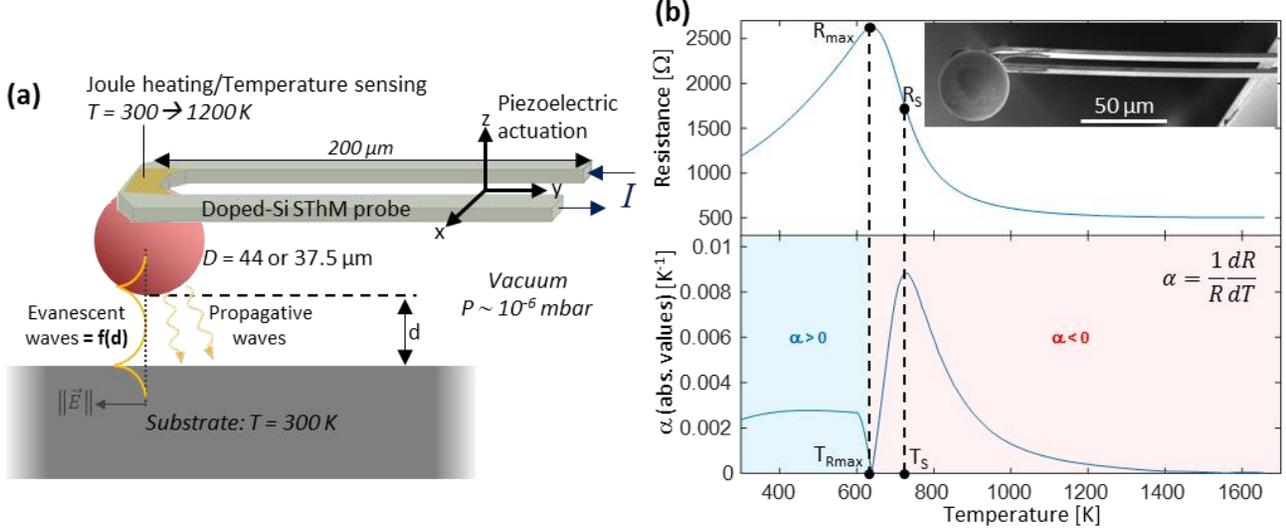

**Fig. 1: Experimental setup.** (a) Schematic of the measurement system composed of a spherical emitter glued on the apex of a doped-silicon SThM probe heated by Joule effect, separated by a gap distance from a bulk substrate. The emitter is displaced by means of piezoelectric actuators. (b) Electrical resistance and temperature coefficient of electrical resistance **α** as a function of the probe apex temperature. The inset is a SEM image of the emitter.

### 2.2. Calibration of the emitter

The temperature dependence of the electrical resistance of the SThM probe $R(T)$ is assessed by a calibration method from room temperature up to 1600 K, from which the temperature coefficient $\alpha = \frac{1}{R}\frac{dR}{dT}$ is obtained (Fig. 1b). This calibration is a key step for determining the link between the electrical resistance and the temperature[38–40]. First, the {probe+sphere} system is placed in an oven usually used for thermocouple calibration (Fluke 9144) and the electrical resistance is measured while the oven temperature is increased from room temperature up to 140 °C. Above this temperature, some components of the holder of the SThM probe such as glue and welding are damaged. Second, the emitter is put under vacuum at room temperature and an electrical current is applied up to 11 mA allowing to measure its electrical resistance as a function of the electrical power as it is heated by Joule effect. A curve with the same shape as that shown in Fig. 1b is measured. In the area where the electrical resistance is decreasing ($T > T_{Rmax}$), the temperature of the emitter is proportional to the electrical power[38]. From room temperature up to a few kelvins below $T_{Rmax}$, the data measured in the oven are well fitted with a quadratic law. Finally, the whole $R(T)$ curve is reconstructed with the quadratic fit of the resistance measured in the oven ($T < T_{Rmax}$) and the temperature deduced from the electrical measurements ($T < T_{Rmax}$). Because the quadratic fit is not valid up to $T_{Rmax}$, putting the two curves together leads to a systematic error of a few percents. In addition, the contribution of the cantilever on the total electrical resistance of the probe can become significant as temperature rises, thus leading to another systematic error. The total uncertainty on the temperature determination is estimated to be about 20 %.

### 2.3. Near-field radiative conductance measurements

The electrical resistance of the SThM probe is part of a Wheatstone bridge where a DC electrical current is supplied by a Keithley 6221 signal generator. The unbalance voltage of the bridge, amplified 100 times, is recorded by a NI cDAQ-9178 data acquisition device with a NI 9239 voltage measurement, unit using an acquisition rate of 2 kHz, while a voltage ramp (from 0 to 100 V) is applied to the z-piezoelectric actuator corresponding to a displacement of 5 µm. The electrical resistance of the SThM probe is deduced from the unbalance voltage of the Wheatstone bridge as a function of the z-piezoelectric actuator displacement. Measurements are repeated 100 times and averaged for enhancing the signal-to-noise ratio, then a sliding average over 40 points is applied in order to reduce the 50 Hz electrical noise. The temperature of the emitter and the temperature coefficient $\alpha$ are inferred from the averaged electrical resistance measurement using the $R(T)$ calibration curve. The near-field radiative thermal conductance $G_{NF}$ is equal to the variation of the total thermal conductance of the emitter $G_{tot} = P/\theta$ as a function of distance, determined from a logarithmic derivation of the thermal conductance:

$$\frac{\Delta G_{tot}}{G_{tot}} = \frac{\Delta P}{P} - \frac{\Delta \theta}{\theta} = \frac{\Delta R}{R} + 2\frac{\Delta I}{I} - \frac{\Delta \theta}{\theta}, \tag{1}$$

with $\theta = T - T_{amb}$ the temperature elevation of the emitter, $R$ the electrical resistance, $P$ the electrical power and $I$ the electrical current. Then, $G_{NF} = \Delta G_{tot}$ can be expressed as a function of the temperature coefficient $\alpha$ as:

$$G_{NF} = G_{tot}\left[\left(T_{ref} - T\right)\left(\frac{1}{\theta} - \alpha\right) + 2\frac{I - I_{ref}}{I}\right], \tag{2}$$

with $T_{ref}$ and $I_{ref}$ the reference temperature and current of the emitter taken at the largest emitter-substrate distance. At the largest distance of ~5 µm, the near-field radiative conductance $G_{NF}$ is equal to 0 because it is a variation of the total thermal conductance compared to the largest distances. Since $G_{NF}$ is representative of the contribution of evanescent waves, assuming it is equal to 0 near 5 µm is a reasonable assumption. The uncertainty of the measurements of $G_{NF}$ corresponds to the sum of the standard deviation and the relative uncertainty of 20 % from the calibration. It was chosen to sum the uncertainties so it corresponds to the worst-case scenario. Finally, a thermal conductance resolution down to ~30 pW.K⁻¹ is found after applying the sliding average.

## 3. Results

### 3.1. Numerical predictions

The temperature dependence of the near-field radiative heat transfer can be determined from the well-known expression of the flux $q(d)$ between two parallel semi-infinite media put forward by Polder and van Hove[1], where $d$ is the distance between the two bodies. As an alternative to the flux, it is also usual to consider the near-field thermal conductance per unit area $g$, which is obtained by dividing $q$ by the temperature difference $\Delta T$ between the two bodies. In order to get the temperature dependence, the flux (or the conductance) can be fitted as a function of temperature to an analytical expression close to that of Stefan-Boltzmann, but where the temperature exponent is considered unknown:

$$q = g(T - T_0) = C\sigma(T^n - T_0^n),\qquad(3)$$

with $T$ and $T_0$ the hot and cold body temperatures, and $\sigma$ the Stefan-Boltzmann constant. The fitting parameters are $C(d)$, the pre-factor related to the emissive properties of the materials, and $n(d)$, the exponent that can differ from 4. These two factors are expected to depend on the distance $d$. Note that $C$ is a function of the monochromatic emissivities of both materials in the far field, and that it can be considered as temperature-independent only for weak temperature differences in principle. Albeit not well studied until now, slight deviations from $n = 4$ could also take place in the far field. In order to analyze the distance dependence, the radiative thermal conductance between two semi-infinite planar materials having a dielectric function $\varepsilon = 1.1 + 0.01\,i$ (arbitrarily chosen, here frequency-independent) is calculated by means of fluctuational electrodynamics as a function of distance. The total conductance is considered, as well as its two evanescent contributions respectively associated with frustrated or surface modes (Fig. 2a). Here, $T_0 = 300$ K and $T$ is ranging from 300 up to 1200 K. Calculations are fitted by the temperature power law (Eq. (3)) for each gap distance. It can be seen that the accuracy of the fits is excellent (dots and lines are superimposed in Fig. 2a, see Fig. A.2 in the supplementary information (Appendix A) for more quantitative analysis). The slope of the conductance as a function of temperature changes when the gap distance varies, which indicates that the exponent of the temperature power law varies too. In Fig. 2b, the exponent is shown as a function of distance. The exponent of the total conductance levels off at large distances, reaching the far-field value $n_{FF} = 4$ and $C = 1$ typical of blackbodies and corresponding to the propagative wave contribution. As the gap distance decreases, the contribution of the evanescent waves to radiative heat transfer increases significantly and causes the exponent to tend to 2 at the smallest distance. This value is caused by the surface modes representing the main contribution to evanescent-wave radiative heat transfer at the smallest distances. Such analysis can be done in principle for all materials and geometries (see section A.1 in the supplementary information for the case of a sphere facing a flat plane), as it depends only on the dielectric function $\varepsilon(\omega)$, where $\omega$ is the angular frequency. It is worth noticing that while the dielectric function is usually well tabulated in the infrared at room temperature, data often lack at higher temperatures.

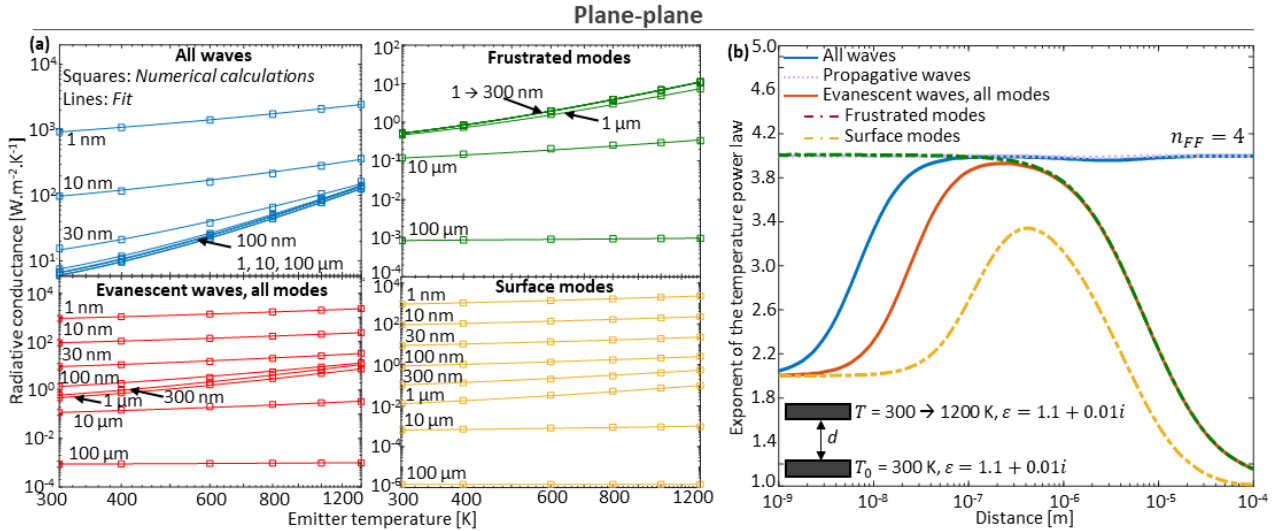

**Fig. 2: Determining the temperature power law of the radiative conductance.** (**a**) Numerical calculations (squares) of the radiative thermal conductance between two planar materials having both a dielectric function $\boldsymbol{\varepsilon} = 1.1+0.01\mathrm{i}$. Numerical results are fitted by the analytical expression (Eq. (3)

of the radiative conductance (lines). (b) Exponent of the temperature power law as a function of distance between the two planar bodies considering the total radiative heat flux (blue, solid), the propagative wave contribution (cyan, dot), the evanescent wave contribution (red, solid) with the frustrated (brown, dash-dot) and surface modes (yellow, dash-dot).

### 3.2. Experimental measurements

We report on measurements of radiative conductance enhancement as the gap distance between a spherical emitter and a planar substrate is decreased from about 5 µm down to a few tens of nanometers. The substrates have dimensions orders of magnitude larger than the size of the sphere, providing a constant sphere-substrate view factor of 0.5 as the gap distance is changed. As a result, no variation of radiative flux is predicted by the macroscopic theory (propagative waves): only near-field thermal radiation is monitored. The spherical shape is chosen because it allows avoiding any parallelism issue encountered with a planar emitter, while its large area still allows for comparisons with theory for flat surfaces (see below). Bulk substrates made either of $SiO_2$, InSb (non-intentionally doped with a residual donor concentration $N_D = 10^{15}$ cm$^{-3}$) or graphite are considered in the following. Measurements could be performed at temperatures larger than the melting temperature of InSb (800 K), because the thermal resistance between the sphere and the substrate into contact, coupled with the low thermal conductivity of $SiO_2$, prevent excessive local heating or melting of InSb.

In general, the last distance before contact is difficult to determine because of mechanical vibrations of the cantilever with respect to the substrate, which are found to have an amplitude of 7 nm using interferometry, and of attraction forces also measured by AFM leading to a snap-in of 2-3 nm (see section A.3). Furthermore, spheres are rough, with a peak-to-peak value of 20-30 nm measured by atomic force microscopy (AFM), which complicates the comparison with theoretical models of perfectly-smooth spheres. As a result of this analysis, the last distance before contact is estimated to be about 30 nm with a $SiO_2$ sphere, and 40 nm with a graphite one. This is in agreement with previous experiments[9,10,41], where the comparison of theoretical and experimental approach curves is performed by matching the data when translating them as a function of distance. Here, we do not shift the z-axis and provide the raw data recorded from the z-piezoelectric actuator ($z = 0$ is the contact), and we underline the distance range where comparison with theory is not allowed (grey-shaded areas in Fig. 3). Nonetheless an analysis with a shifted z-axis is proposed in the supplementary information (Fig. A.7).

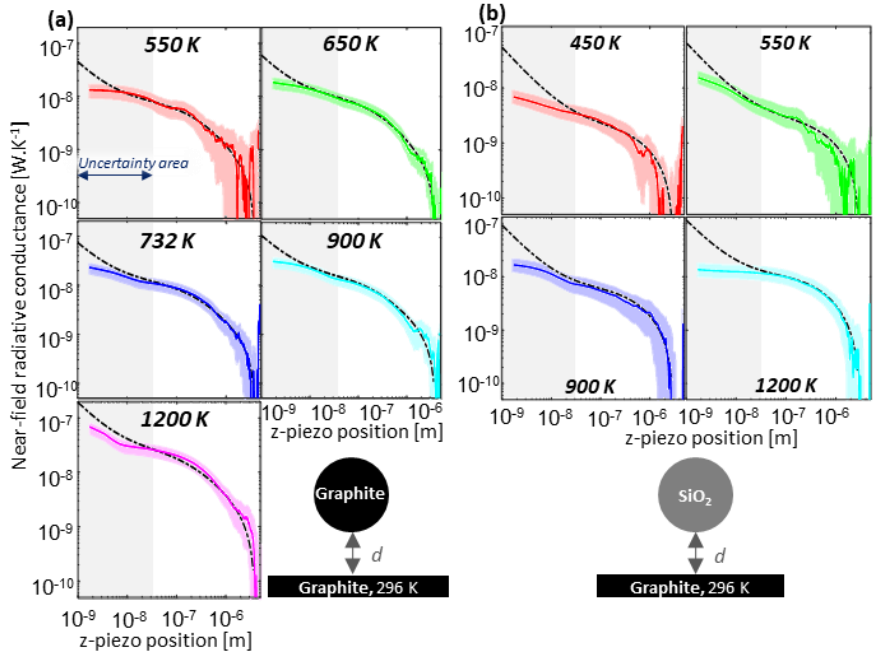

**Fig. 3: Near-field radiative conductance measurements between a spherical emitter made of graphite or silica and a graphite substrate as a function of z-piezoactuator position and emitter temperature**. (a) Graphite emitter (symmetric case). (b) Silica emitter. The black dash-dotted line represents calculations from the proximity approximation. The grey-shaded area represents the range where there are large distance determination uncertainties induced by the roughness of the materials and mechanical vibrations. Note that the radii of the graphite (18.8 μm) and silica (22 μm) spheres are slightly different.

Fig. 3 provides an example of measurements for a graphite substrate as a function of distance and temperature, involving a symmetric case (graphite sphere) and a non-symmetric case (silica sphere). Thermal conductances $G$ are given in W.K$^{-1}$. It is seen that for the graphite-graphite configuration thermal radiation transfer is efficient, reaching a record near-field thermal conductance of 68.9 ± 13.7 nW.K$^{-1}$ due to the large emitter temperature. However, the ratio $G_{max}(T = 1200$ K$)/G_{max}(T = 550$ K$)$ reaches hardly 5, while a fourth-power dependence predicts about 23. This difference is found in spite of graphite being often considered as a paradigmatic material as close as possible to the blackbody.

The experimental data are systematically compared with theoretical results obtained using the proximity (also called Derjaguin[42]) approximation (referred to as PA) which accounts for the evanescent wave contribution only (see section A.1 for an example with $\varepsilon = 1.1 + 0.01\ i$). Temperature-dependent dielectric functions were considered for SiO$_2$ and InSb, as they are known respectively from Joulain et al.[43] and Vaillon et al.[44], but not for graphite. For the graphite-graphite case the measured conductances seem to agree with calculations for distances down to ~10 nm, which may indicate that the last distance before contact is smaller than that expected for this configuration. This may be caused by the relative softness of graphite compared to the two other investigated materials, which may flatten the contact area and strongly reduce the influence of roughness, leaving only mechanical vibrations and attraction forces as impacts on the last distance before contact. Note that this distance is similar to the one found by Sahiloglu et al[45].

We have systematically investigated the radiative transfer between $SiO_2$ and graphite hot spherical emitters, and $SiO_2$, InSb and graphite substrates at room temperature. All approach curves are provided in the supplementary information. Among the observed peculiarities, we find that the exchanged thermal radiation in the $SiO_2$-$SiO_2$ configuration is much weaker than that predicted with the PA for symmetric $SiO_2$ bodies. We verified that the dielectric function of the used $SiO_2$ sphere (supplied by Corpuscular Inc.) is different from the usual tabulated $SiO_2$ bulk one (see section A.3). Note that it is not the first time that such a deviation is observed[27]. Since spectral matching is of paramount importance for materials supporting phonon-polaritons, it is logical that the flux is weaker if the material of the sphere is slightly different from that of the substrate. The exchanged thermal radiation between the $SiO_2$ sphere (therefore termed modified $SiO_2$ in the following) and the other substrate materials (InSb, graphite) is much closer to the predictions of the PA, highlighting the fact that the exact composition, and very likely the position of the phonon-polariton peaks, is less significant for the near-field radiative exchange between asymmetric configurations. In particular, InSb phonon-polariton peaks are located at larger wavelengths (~55 µm) than the thermally-excited spectral range and do not contribute significantly to the near-field radiative heat transfer with the two types of spheres.

**Temperature dependence**

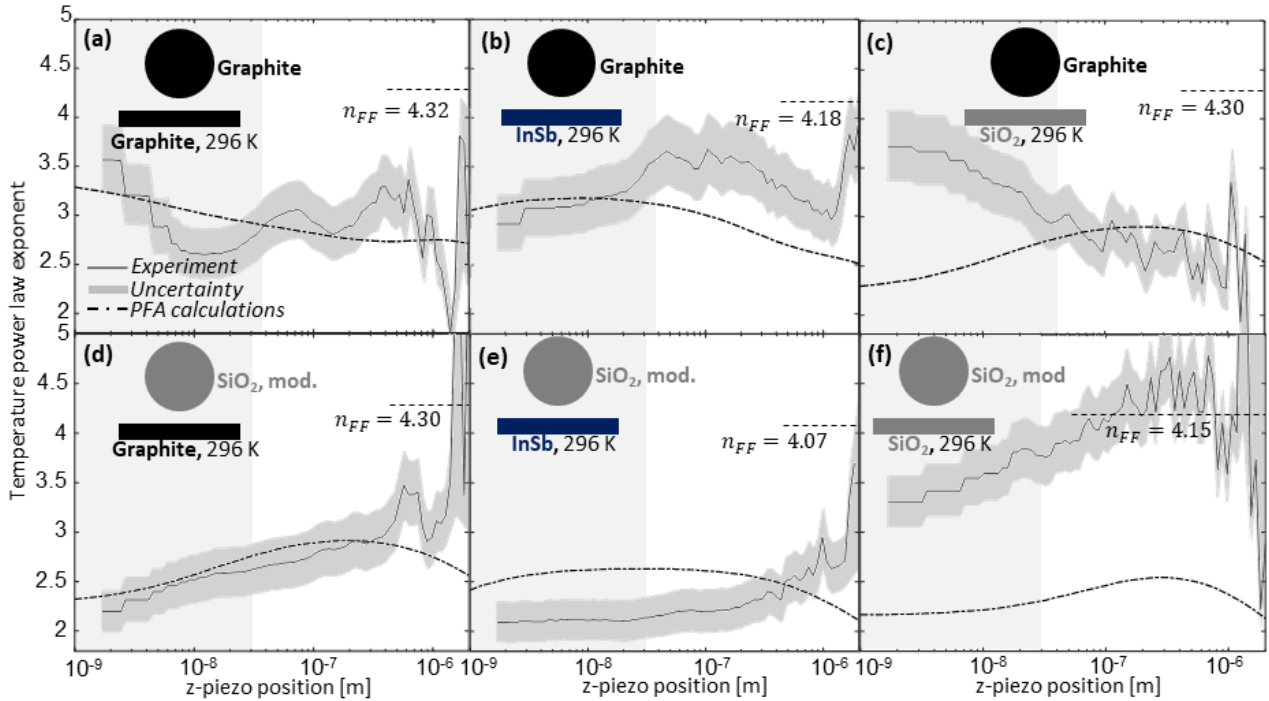

**Fig. 4: Exponent of the temperature power law of the near-field radiative thermal conductance as a function of z-piezo position.** (a) graphite-graphite, (b) graphite-InSb, (c) graphite-$SiO_2$, (d) Modified $SiO_2$-graphite, (e) Modified $SiO_2$-InSb, (f) Modified $SiO_2$-$SiO_2$. Grey lines are for the experimentally-determined exponent, the grey areas represent the uncertainty and the black dash-dotted lines are calculations from the proximity approximation.

Temperature dependence is investigated for each pair of materials. The exponents of the temperature power law of the near-field radiative conductance are deduced from measurements fitted by Eq. ((3) and are quantitatively compared to those obtained from PA calculations. Results are shown in Fig. 4 as a function of distance. First, note that the exponent $n_{FF}$ calculated in the far field is shown in order to highlight the differences with the measurements performed in the near field accounting for the evanescent waves. Values of $n_{FF}$ are found above the value of 4 of the two-blackbody case because emissivity of the materials depends on both frequency and temperature. When temperature increases, the radiative heat flux in the far field for these materials is enhanced at a faster apparent rate than that between two blackbodies: this is due to the increasing monochromatic emissivity of the materials at high frequency (see section A.6 and Fig. A.9). Second, the exponent in the fitted law (Eq. (3) is smaller for the near-field contribution. As distance decreases, the exponent resulting of the contribution of evanescent waves increases to a maximum value near a distance of 1 µm then decreases to 2. This changing behavior is due to the increasing contribution of the surface modes compared with the frustrated modes.

Different behaviors are expected in the near field depending on materials. For the graphite-graphite configuration (Fig. 4a) the exponent keeps increasing as distance decreases. This may be due to the fact that graphite does not support any surface polariton, unlike the two other materials, so the main contribution of the surface modes is not expected to be localized near a single frequency. The general expected evolution of the exponent is observed experimentally. For the configurations with an InSb substrate (Fig. 4b,e) the exponent is expected to have an almost flat behavior below 100 nm because the contribution of surface modes does not increase sufficiently for impacting the temperature power law of the near-field radiative conductance. The expected behavior of the exponent is observed experimentally but without a quantitative agreement with calculations. For the modified $SiO_2$-graphite (Fig. 4d) and graphite-$SiO_2$ (Fig. 4c) configurations, the exponent is expected to reach a maximum value near 300 nm then tend to 2.2-2.3 as distance decreases. A relatively good agreement is found below 1 µm for the modified $SiO_2$-graphite configuration and for distances ranging between 30 nm and 2 µm for the graphite-$SiO_2$ configuration. For this last configuration, the large deviations of the experimentally determined exponents in the small-distance range compared to calculations could be explained by the levelling off of the measurements (Fig. 3a) due to distance uncertainties. Finally, the modified $SiO_2$-$SiO_2$ (Fig. 4f) experimental and numerical data do not match, which is in agreement with the fact that they do not match either as a function of distance (the dielectric function for the sphere is different from that in tabulated data).

We remind that the temperature dependence of the permittivity is known for $SiO_2$ but not for graphite (see section A.5 in the supplementary information). For materials whose permittivity is known at room temperature, it may be possible to deduce some information on the temperature dependence of the permittivity of the spheres based on the differences between the predicted and the measured exponents.

## 4. Discussion

| Sphere-substrate configuration | $\Delta T_{max}$ [K] | Maximum conductance at $\Delta T_{max}$ [nW.K⁻¹] | Temperature power law exponent of the near-field radiative conductance | | |
|---|---|---|---|---|---|
| | | | Far-field (calculated) | Calculated | Measured |
| | | | | $d$ = 100 nm | |
| Graphite-SiO₂ | 477 | 4.9 ± 1.0 | 4.30 | 2.88 | 2.84 ± 0.31 |
| Modified SiO₂-SiO₂ | 493 | 7.4 ± 1.5 | 4.15 | 2.46 | 4.11 ± 0.33 |
| Modified SiO₂-InSb | 904 | 7.6 ± 2.1 | 4.07 | 2.61 | 2.21 ± 0.22 |
| Graphite-InSb | 448 | 10.8 ± 2.1 | 4.18 | 3.01 | 3.67 ± 0.34 |
| Modified SiO₂-Graphite | 904 | 16.7 ± 3.3 (at $\Delta T$ = 604 K) | 4.30 | 2.89 | 2.80 ± 0.27 |
| Graphite-Graphite | 904 | 68.9 ± 13.7 | 4.32 | 2.82 | 2.92 ± 0.29 |

**Tab. 1:** Summary of the main results reported from the study of the near-field radiative conductance as a function of distance and emitter temperature for different pairs of materials. The different configurations are ordered by their maximum near-field radiative conductance. The uncertainty considers the sum of the standard deviation of the conductance measurements and the systematic error of the temperature measurement due to the calibration.

A summary of the different studied configurations is shown in Tab. 1, reporting on the maximum sphere-substrate temperature difference, the maximum near-field radiative conductance and the measured and calculated exponents of the temperature power law at a distance of 100 nm. This distance is selected because it is larger than the uncertainty range while still representing the near field. The results are ordered by their maximum measured conductance. In general, larger conductances are measured for larger temperature differences except with the substrate made of InSb. This is due to the fact that the spectral matching with the optical properties of InSb is better for graphite than with that of modified SiO₂, thus leading to a larger enhancement of radiative heat transfer due to the contribution of evanescent waves. The exponent of the temperature power law measured at 100 nm is ranging from 2.21 for the modified SiO₂-InSb configuration up to 4.11 for the modified SiO₂-SiO₂ configuration, demonstrating a strong material dependence of near-field radiative heat transfer. As expected by calculations, very similar exponents are found for the two inverse configurations with 2.84 ± 0.31 for graphite-SiO₂ and 2.80 ± 0.27 for modified SiO₂-graphite.

## 5. Conclusion

In conclusion, we have reported on measurements of the near-field radiative conductance as a function of distance in the last 5 micrometers between a sphere made of SiO₂ or graphite and a planar substrate made of either SiO₂, graphite or InSb. The sphere was heated up to 1200 K providing large temperature differences with the substrate up to 900 K. A large near-field radiative conductance close to 70 nW.K⁻¹ was measured for the graphite-graphite configuration with a gap distance of ~40 nm. We have analyzed the temperature dependence of near-field radiative heat transfer by determining the exponent of the temperature power law of the near-field radiative conductance as a function of distance for six pairs of materials. Temperature power laws in the near field have been found very different – generally smaller – from those expected in the far field, with significant differences from one material pair to another. Similar experiments should be performed with other geometries, such as planar and tip-shaped emitters or micro-structured and multilayer substrates, in order to investigate the geometrical dependence of the near-field radiative conductance as a function of temperature. Furthermore, high-temperature near-field radiative heat transfer matters in energy harvesting applications, such as thermophotovoltaics[39], in order to significantly enhance their output power densities . However, these results highlight that increasing the emitter temperature in near-field thermophotovoltaic devices will unfortunately not amplify the converted power as much as they do in the far field. It is therefore critical for such applications to bring the emitter and the cell as close as possible.

## Author contributions

R.V and P.-O.C conceived and supervised the work. C.L fabricated the emitter and performed the experiments. C.L and R.V performed the simulations. The manuscript was written by all authors.

## Declaration of competing interest

The author(s) declare no competing interests.

## Acknowledgment

The authors thank D. Cakiroglu, J.P. Perez, T. Taliercio and E. Tournié for useful discussions and for providing InSb samples. C.L and P.-O.C thank C. Ducat, N. Pouchot, X. Durand and A. Buthod for their assistance in the design of the experimental setup, D. Renahy, P. Mangel, E. Guen and S. Gomès for experimental support, M. Piqueras and J.M. Bluet for data comparisons, and A. Delmas for FTIR measurements. P.-O.C further thanks S. Volz for equipment handover. The authors acknowledge support by the French National Research Agency (ANR) (ANR-16-CE05-0013, DEMO-NFR-TPV project), partial funding by the French "Investment for the Future" program (EquipEx EXTRA ANR-11-EQPX-0016 and IDEXLYON ANR-16-IDEX-0005) and by the Occitanie region, and EU project EFINED (H2020-FETOPEN-2016-2017 GA 766853).

## Appendix A. Supplementary information

Supplementary data to this article can be found online at XXX

# Supplementary information

## Temperature dependence of near-field radiative heat transfer above room temperature

C. Lucchesi[1,2]*, R. Vaillon[2,1], P.-O. Chapuis[1]

[1] Univ Lyon, CNRS, INSA-Lyon, Université Claude Bernard Lyon 1, CETHIL UMR5008, F-69621, Villeurbanne, France
[2] IES, Univ Montpellier, CNRS, Montpellier, France

christophe.lucchesi@insa-lyon.fr

## A.1. Predictions of thermal radiative conductances

An analysis similar to that of Fig. 2 is performed with a spherical emitter instead of a planar one facing a planar substrate. The Proximity Approximation (Derjaguin) is applied for the evanescent waves. In Fig. A.1a, the total conductance as a function of emitter temperature is almost independent of distance because of the small contribution of evanescent waves caused by the curved geometry of the emitter. In fact, the curvature of the surface of the sphere leads to a variable distance gap with the surface of the sample, ranging from $d$ at the closest and up to $d + R$ at the largest, $R$ being the radius of the sphere. For a sphere having a radius of the order of a few tens of micrometers (20 µm in Fig. A.1a,b), a major part of the sphere surface is too far from that of the planar material to allow evanescent waves to contribute significantly to radiative heat transfer. At the largest distance, both the plane-plane and sphere plane geometry tend to the same total radiative conductance (compare Fig 1a of the main paper and Fig A.1a) because the effect of the curvature of the sphere becomes negligible. Another effect of the curvature of the sphere is a flattening of the exponent for the different contributions in the near field while similar exponents are found in the far field compared with a planar emitter (Fig A.1b).

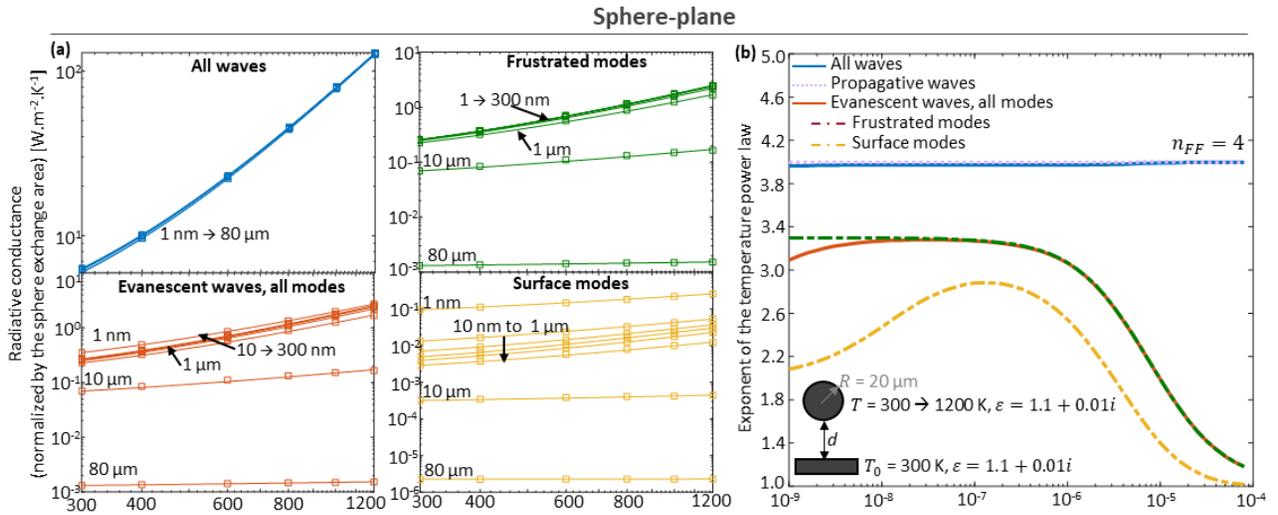

**Fig. A.1: Determining the temperature power law of the radiative conductance between a sphere and a plane.** (**a**) Numerical calculations (squares) of the radiative thermal conductance for the media having both a dielectric function $\varepsilon = 1.1 + 0.01i$. Numerical results are fitted by the analytical expression (Eq. 3**Erreur ! Source du renvoi introuvable.** in the manuscript) of the radiative conductance (lines). (**b**) Exponent of the temperature power law as a function of distance between materials considering the total radiative heat flux (blue, solid), the propagative wave contribution (cyan, dot), the evanescent wave contribution (red, solid) with the frustrated (brown, dash-dot) and surface modes (yellow, dash-dot).

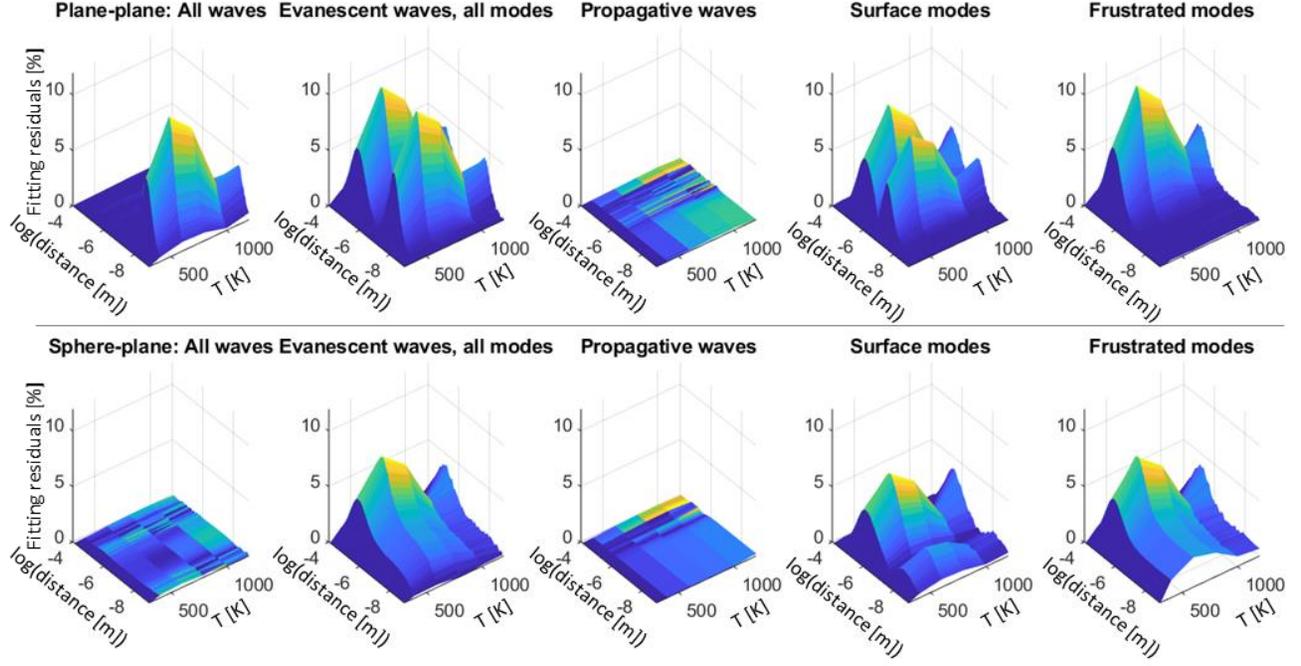

**Fig. A.2: Relative residuals (fitting deviations) of the fits of the numerical calculations by the analytical temperature power law (Eq. 3) as a function of distance (logarithmic scale) and emitter temperature.** Residuals are shown for each wave contributions and for both the plane-plane (top) and the sphere-plane (bottom) configurations.

Fig. A.2 shows 2D-plots representing the relative residuals of the fit of the fluxes from numerical calculations (fluctuational electrodynamics) by the analytical power law (Eq. 3) as a function of distance and emitter temperature, expressed as:

$$residuals(d,T)[\%] = 100 \times \frac{\left|q_{numerical}(d,T) - q_{fit,analytical}(d,T)\right|}{q_{numerical}(d,T)}. \qquad (A.1)$$

It appears that the fits are very good because relative residuals are never higher than 10 % of the numerical calculations.

## A.2. Near-field radiative heat transfer measurements

We provide here all the data of the near-field radiative heat transfer measurements.

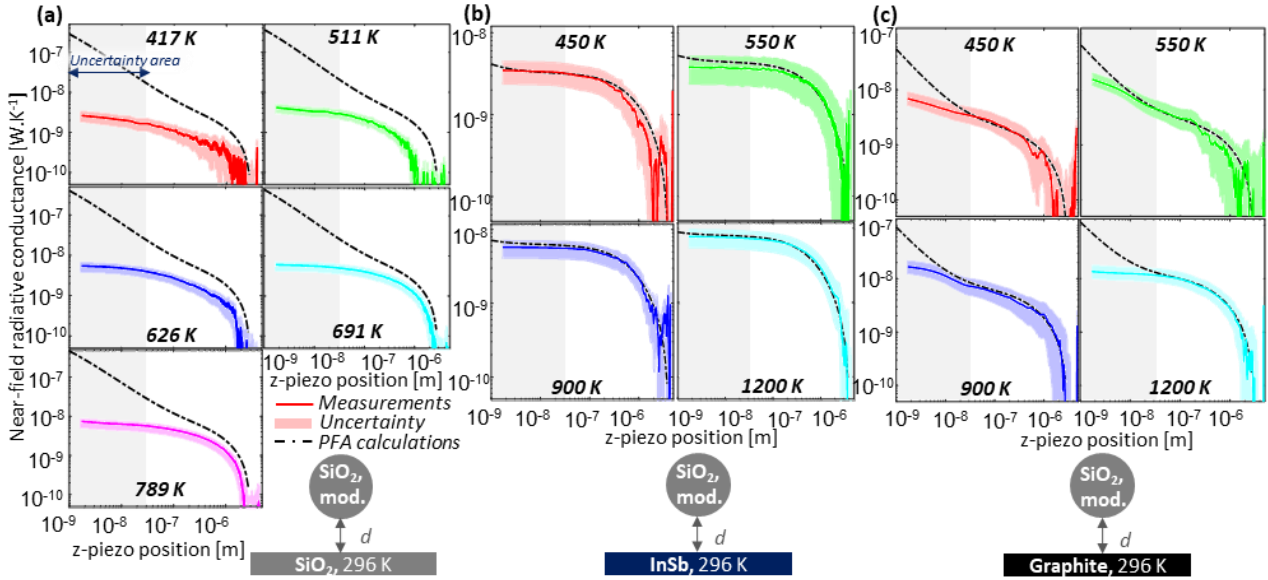

**Fig. A.3: Near-field radiative conductance measurements with a modified SiO₂ emitter.** Near-field radiative conductance between a modified SiO₂ sphere heated from 417 to 1200 K and a planar substrate at room temperature made of either SiO₂ (**a**), InSb (**b**) or graphite (**c**), as a function of z-piezo position. The grey-shaded area represents the range where there are large distance determination uncertainties induced by the roughness of the materials and mechanical vibrations (see Fig. A.6).

In all configurations a good agreement is found between measurements and calculations for distances above 30 nm, except for the symmetrical case with the emitter and the substrate both made of SiO₂ (Fig. A.3a). This disagreement may be explained by a frequency shift of the dielectric function observed by reflectivity measurements on the sphere compared to that of the bulk substrate, which may significantly affect radiative heat transfer (see section A.3). The sphere permittivity is therefore termed 'modified SiO₂'.

The largest near-field radiative conductance of $16.7 \pm 3.3$ nW.K$^{-1}$ is found for the modified SiO₂-graphite configuration (Fig. A.4c) with the sphere at 900 K, which is larger than the maximum value measured at 1200 K. This unexpected result is explained by the last distance before contact (driven by roughness and vibrations, see section A.4) that might be smaller for the experiment performed at 900 K, thus leading to a near-field radiative conductance larger than that measured at 1200 K.

For the modified SiO₂-InSb configuration (Fig. A.3b), both measurements and calculations level off at low distances because the dielectric functions of the two materials are not matching well in the frequency range where most of the radiative heat transfer occurs for temperatures ranging from 450 to 1200 K (see Fig. A.8a,b).

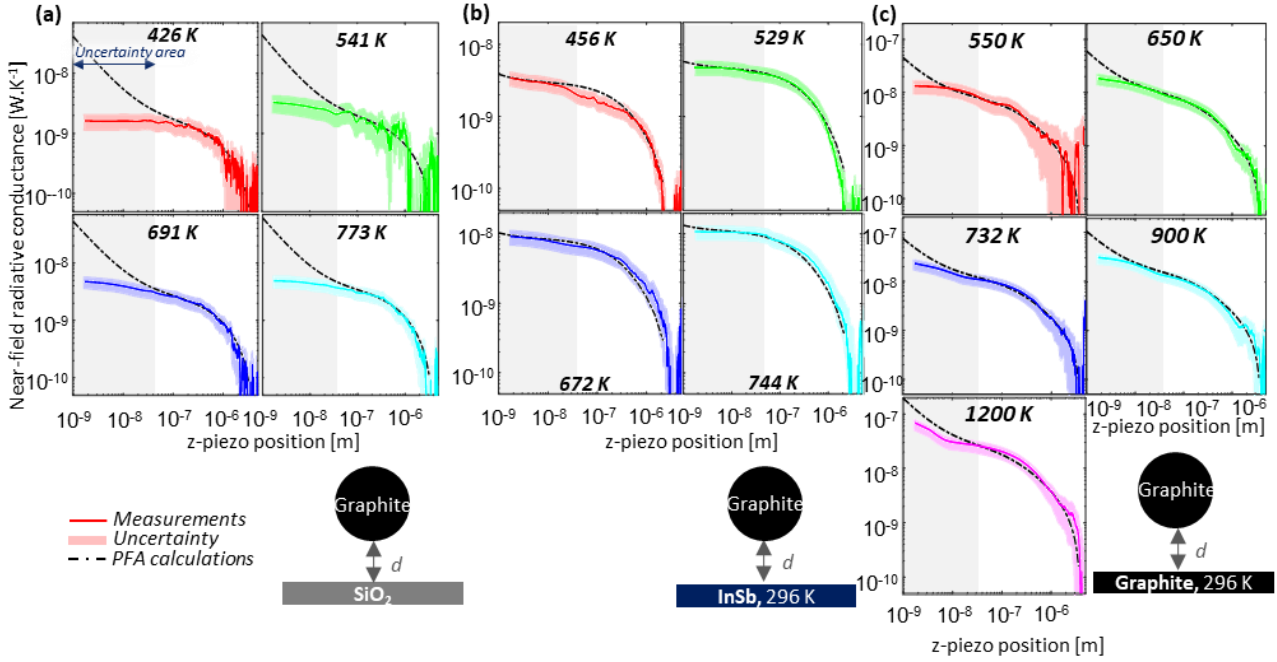

**Fig. A.4: Near-field radiative conductance measurements with a graphite emitter.** Near-field radiative conductance between a graphite sphere heated from 426 to 1200 K and a planar substrate at room temperature made of either $SiO_2$ (**a**), InSb (**b**) or graphite (**c**), as a function of z-piezo position. The grey-shaded area represents the range where there are large distance determination uncertainties induced by the roughness of the materials and mechanical vibrations.

If the $SiO_2$ sphere had the same permittivity as the substrate, the graphite-$SiO_2$ configuration (Fig. A.4a) would be the opposite in terms of materials compared to the $SiO_2$-graphite configuration previously studied. According to calculations, thermal rectification may be observed between the two configurations because of the temperature dependence of the dielectric function of $SiO_2$, which was measured in the work of Joulain *et al.*[1]. In this case, conductance differences up to a few percent may be expected. Unfortunately, the large distance uncertainties close to contact, the accuracy of the conductance measurement and the permittivity variation could not allow us to conclude on an observation of thermal rectification.

For the graphite-InSb configuration (Fig. A.4b) the temperature of the sphere is kept below the melting temperature of InSb because the thermal conductivity of graphite (25-470 W.m.$^{-1}$K$^{-1}$) is one to two orders of magnitude larger than that of $SiO_2$ (1.4 W.m.$^{-1}$K$^{-1}$). A contact between a graphite sphere heated above 800 K and an InSb substrate may damage the sample and pollute the sphere.

The symmetrical graphite-graphite configuration provides the largest conductance among all configurations studied in this work, with a maximum of 68.9 ± 13.7 nW.K$^{-1}$ measured at an emitter temperature of 1200 K ($\Delta T$ = 904 K).

### A.3. Silica spheres

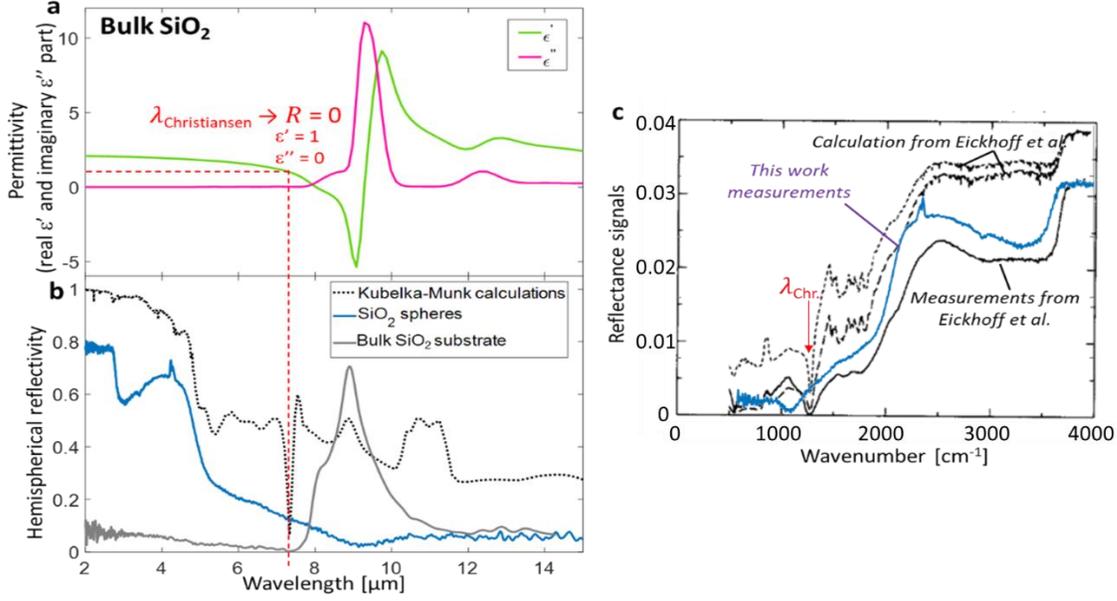

**Fig. A.5: Reflectivity measurements on SiO₂ spheres compared to a bulk SiO₂ substrate.** (**a**) Real and imaginary parts of the dielectric function of SiO₂ as a function of wavelength. The Christiansen wavelength where the reflectivity of the material is equal to 0 ($\varepsilon' = 1$ and $\varepsilon'' = 0$) is highlighted. (**b**) Kubelka-Munk reflectivity calculations for the SiO₂ spheres compared to reflectivity measurements for the SiO₂ spheres and a bulk substrate. (**c**) Comparison of this work measurements with measurements and calculations from Eickhoff et al.[2].

The dielectric function of bulk SiO₂ measured by Joulain et al.[1] is shown in Fig. A.5a. SiO₂ has a Christiansen wavelength $\lambda_{Chr}$, where the refractive index of the material is the same as that of its environment (air in our case). Scattering nearly vanishes and almost all the light is transmitted if the material is not absorbing at $\lambda_{Chr}$, leading to a reflectivity that tends to 0. The zero-reflectivity value is measured at the expected wavelength for a bulk SiO₂ substrate (Fig. A.5b) and for the estimated reflectivity of the spheres calculated with the Kubelka-Munk theory[3]. However, $\lambda_{Chr}$ seems to appear at a shifted wavelength for a sample made of SiO₂ spheres, meaning that their dielectric function is different from that of the bulk SiO₂. In Fig. A.5c, reflectivity measurements from this work are compared with similar measurements from the literature performed by Eickhoff et al.[2] with 4-40 µm in diameter SiO₂ spheres. The general behavior is similar between the two sets of measurements but that of Eickhoff exhibits a zero-reflectivity measurement at the expected Christiansen wavelength, contrary to the measurements of this work where $\lambda_{Chr}$ appears again with a shift. The comparison of $\lambda_{Chr}$ between a bulk SiO₂ substrate, the spheres used during this work and measurements from literature, allows to conclude that the dielectric function of this work's SiO₂ spheres is different from that of the bulk substrate and may explain the disagreement between the near-field radiative heat transfer measurements and calculations for the symmetrical SiO₂-SiO₂ case.

For this configuration, surface phonon polaritons (SPhPs) supported by the sphere and the substrate are expected to have the same frequency, thus enhancing drastically radiative heat transfer in the near field. A frequency shift of the dielectric function for the sphere may lead to a non-matching of the SPhPs frequencies between the emitter and the substrate. For configurations with a substrate made of another material than SiO₂, no match of SPhPs frequencies is expected, so that a dielectric function of the $SiO_2$ sphere slightly different than expected should not have any significant impact on near-field radiative heat transfer.

## A.4.  Distance determination

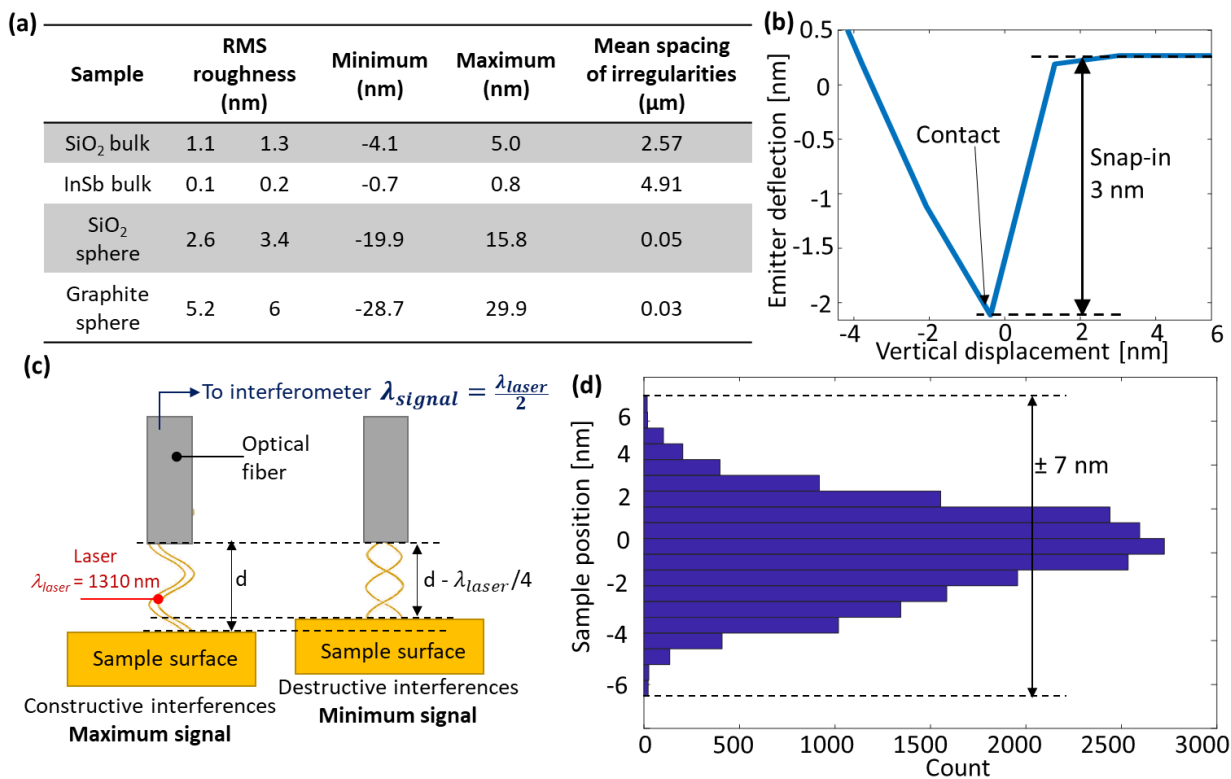

**Fig. A.6: Determination of the distance uncertainty close to contact.** (a) Roughness measurements performed with atomic force microscopy (AFM) on bulk samples and on the SiO₂ and graphite spheres. (b) Emitter deflection measurement as a function of its vertical displacement measured with an AFM setup. (c) Interferometric measurement setup of the mechanical vibrations of the sample. (d) Histogram of the position of the sample around its mean position, measured with the interferometric setup

The distance uncertainty at which the contact between the spherical emitter and the substrate occurs was estimated based on roughness, mechanical vibration and snap-in measurements. The roughness of bulk samples and spheres was measured by atomic force microscopy (AFM). The table in Fig. A.6a summarizes the root-mean-square (RMS) roughness measured on two different sets of samples for each case, the minimum and maximum height of irregularities, and the mean spacing between irregularities. The bulk substrates are very flat with irregularities having a maximum height of a few nanometers. However, the roughness of the spheres is more important with irregularity heights up to 30 nm and more closely spaced compared to those of bulk substrates.

The snap-in of the emitter close to contact was also measured using an AFM setup. Here a laser beam illuminates the cantilever of the SThM probe (where the sphere is attached) and the reflected beam is collected by a quadrant photodiode. When the cantilever bends, a deflection signal is measured and assumed to be proportional to the amplitude of the bending. Close to contact, attraction forces between the spherical emitter and the substrate can bend the cantilever and bring the sphere into contact with the substrate (snap-in). In Fig. A.6b, a snap-in distance of 3 nm is measured, and thus contributes very little to the distance uncertainty compared with the roughness of the spheres.

Vibrations of the setup were measured using an interferometric unit with an optical fiber (Fig. A.6b) by illuminating the surface of the sample with a laser beam having a wavelength of 1310 nm. The laser beam is reflected on the sample and collected back by the fiber and sent to the interferometric unit. The resulting signal, having a wavelength equal to half that of the laser, is used to determine the oscillations of the sample around its mean position. A histogram of the positions of the sample is provided in Fig. A.6c. It appears that the sample oscillates around its mean position with an amplitude of 7 nm.

Adding the contributions of the roughness, the snap-in and that of the vibrations leads to a distance uncertainty of 30 to 40 nm depending on the material of the sphere (the graphite spheres have a larger roughness than modified silica ones).

An estimation of the minimum distance $d_{min}$ reached during the radiative heat transfer experiments can be made by applying a distance shift to the measurements in order to best fit to the PA calculations. To obtain the value of $d_{min}$, a series of distance shifts is applied to each measurement in order to minimize the root-mean-square error between the measured and the calculated distances below 300 nm. The shifted measurements and PA calculations are represented in Fig. A.7 for each pair of materials. Except for the SiO$_2$-SiO$_2$ case (see section A.3), a good agreement is found between the shifted measurements and calculations. The average estimated $d_{min}$ are listed in each sub-figure and are ranging from 6 up to 56 nm (excluding the SiO$_2$-SiO$_2$ case), which is an agreement with the distance uncertainty range that was found with the roughness, snap-in and vibration analysis.

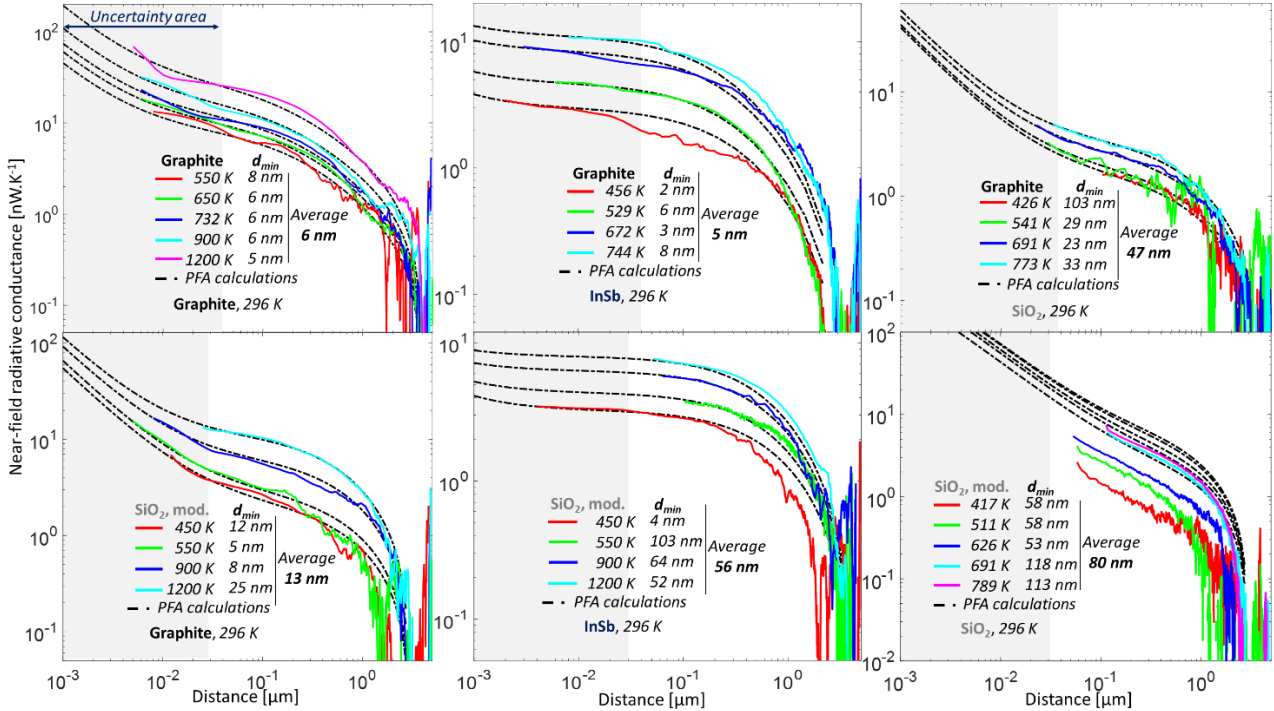

**Fig. A.7: Experimental estimation of the minimum distance before contact.** Near-field radiative conductance calculations using PFA are compared with measurements after fitting and adjusting the distance scale.

## A.5. Dielectric functions

This section provides the real and imaginary parts of the dielectric functions of each material. Data for SiO$_2$ and graphite come from measurements performed respectively by Joulain et al.[1] (from room temperature up to 1480 K, measurements at 295 K are represented in Fig. A.8a) and Querry[4] (data at room temperature only). In the case of InSb the dielectric function was calculated by Vaillon et al.[5].

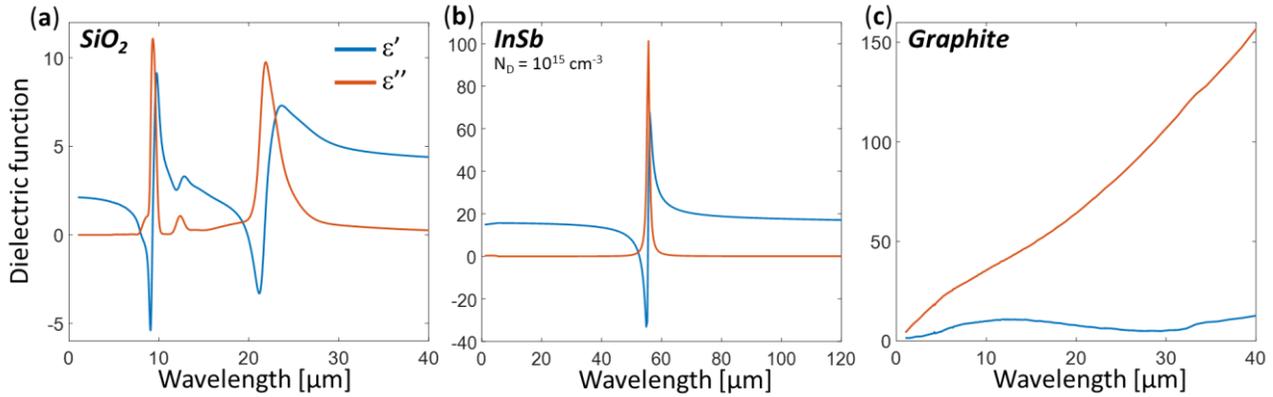

**Fig. A.8: Dielectric function at room temperature as a function of wavelength for the different materials.** (**a**) SiO$_2$ from Joulain et al.[1], (**b**) InSb from Vaillon et al.[5] and (**c**) graphite from Querry[4].

### A.6. Near-field radiative heat transfer calculations for the plane-plane configuration

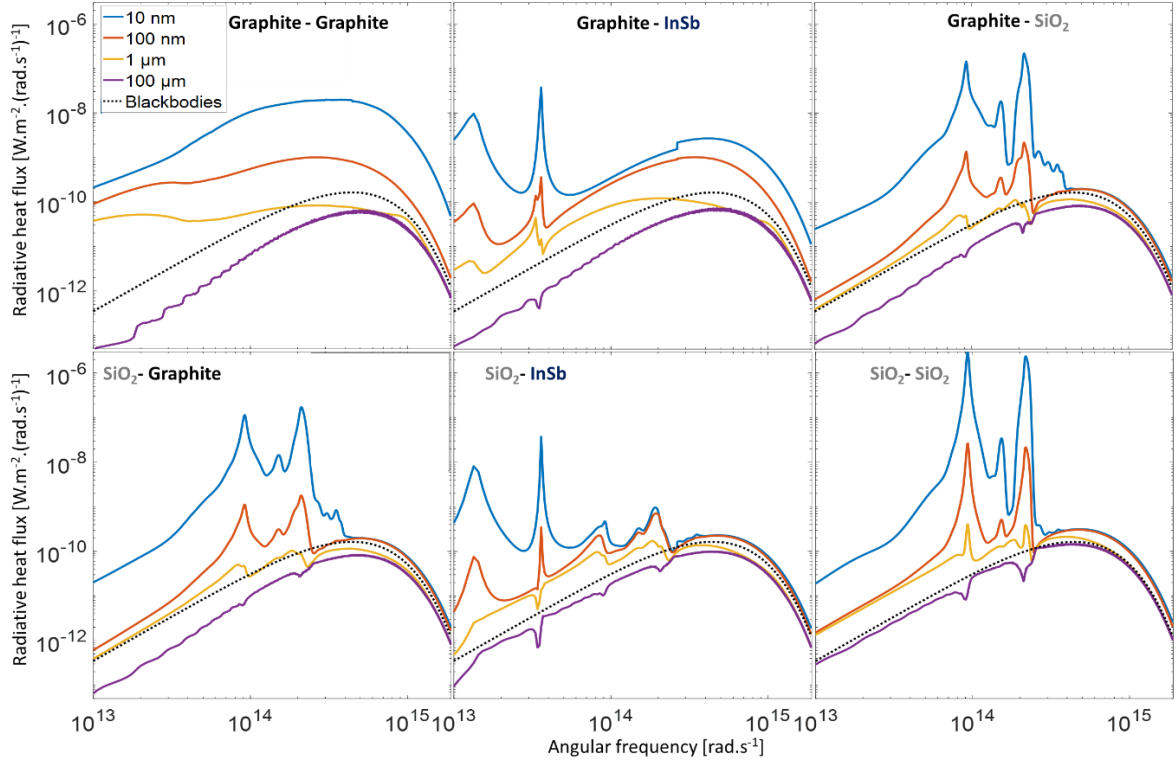

**Fig. A.9:** Radiative heat flux as a function of angular frequency between planar bodies at 1200 K, made of graphite or SiO₂, and planar bodies at 300 K made of either graphite, InSb or SiO₂.

Fig. A.9 represents the radiative heat flux between a planar body made of graphite or SiO₂ at 1200 K and planar bodies at 300 K made either of graphite, InSb or SiO₂. The figure shows the spectra calculated at distances of 100 μm, 1 μm, 100 nm and 10 nm compared to the radiative flux exchanged between two blackbodies. The radiative flux is mainly enhanced at low frequencies but with peaks appearing at the resonance frequencies of the surface polaritons of SiO₂ and InSb. In the far field the fitted flux using the temperature power law (Eq. 3 in the main text) leads to an exponent $n_{FF}$ exceeding the value of 4 (see Fig. 4) representative of the blackbody. This is due to the shape of the spectra of the radiative heat flux for these materials in the far field (curve at d = 100 μm in Fig. A.9). These spectra are lower in amplitude than that between two blackbodies with larger differences at low frequencies meaning a lower emissivity, well seen for the SiO₂-graphite configuration. When temperature increases, the frequency of the maximum radiative heat flux $\omega_{Wien}$ is shifted towards high frequencies (from 1.1 10¹⁴ at 300 K to 4.4 10¹⁴ at 1200 K), where the shape of the spectra for the real materials are close to that of the blackbodies. Therefore, the radiative heat flux in the far field for these materials is enhanced at a faster apparent rate than that between two blackbodies, because of the increasing emissivity of the material at high frequencies.

## A.7. Calibration of the emitter

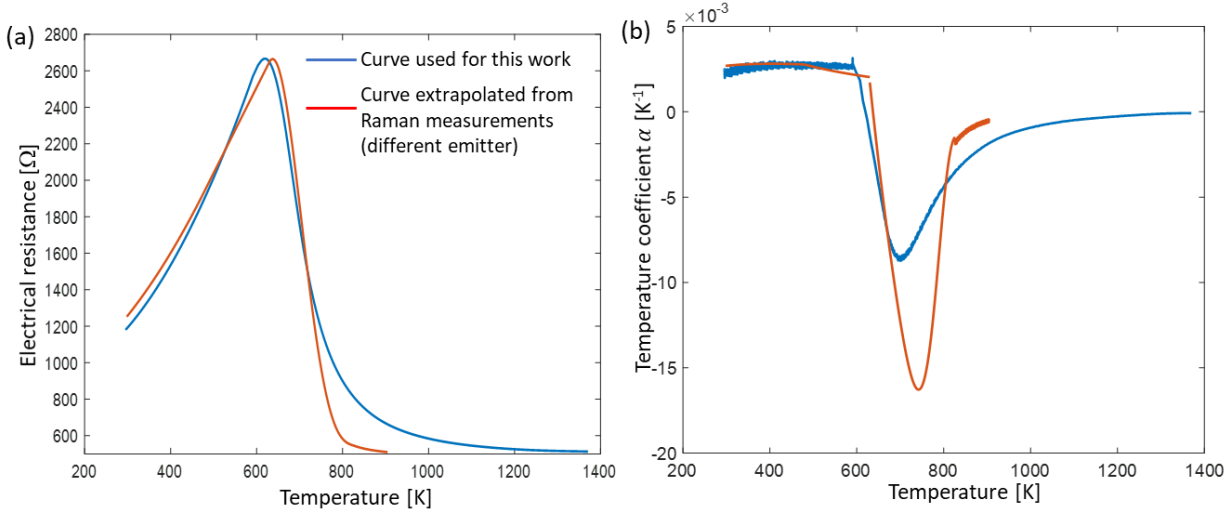

**Fig. A.10: Comparison of two emitter calibration methods.** (a) Electrical resistance as a function of temperature. The blue curve shows the calibration curve used for this work and obtained using the method described in section A.2. The red curve corresponds to a calibration curve extrapolated from Raman temperature measurements performed on another emitter. (b) Temperature coefficients calculated from the two calibration curves.

In order to verify the accuracy of the calibration method described in section A.2, Raman temperature measurements were performed on an emitter while heating it with an electrical current and measuring its electrical resistance at the same time (Fig. A.10a). It is important to remark that the two curves correspond to two different emitters that may have significantly different behaviors, as they are based on SThM probes whose properties strongly depends on fabrication process parameters (doping level). The two curves are very similar up to 750 K but deviate strongly at higher temperatures up to a relative difference of 25 % at 1200 K. This difference is slightly higher than the uncertainty of 20 % considered in this work. Concerning the temperature coefficient $\alpha$ (Fig. A.10b), differences up to a factor of 2 are observed between the two curves. However, possible errors on $\alpha$ have a limited impact on the calculation of the near-field radiative thermal conductance (Eq. 2 in the main text) because in the equation, $\alpha$ is multiplied by a term (temperature difference) that depends on the inverse of $\alpha$. Considering the calibration curve extrapolated from Raman measurements, calculations of near-field radiative conductance and exponent of the temperature power law respectively led to differences of 10 and 15 % respectively.

To summarize, emitter temperatures larger than 750 K that have been measured during this work could have been overestimated by a factor up to 25 % (reached at 1200 K), leading to a conductance overestimated by up to 10 % and an exponent underestimated by up to 15 %. Except for measurements at 1200 K, these potential errors on temperature, conductance and exponent are smaller than the 20 % uncertainty already considered. Near-field radiative heat transfer experiments using Raman-calibrated emitters may help in order to improve the accuracy of the measurements in the future[6].